\documentstyle[pra,aps,preprint]{revtex}
\begin{document}
\title{A general framework of nondistortion quantum interrogation}
\author{Zheng-Wei Zhou$^1$, Xingxiang Zhou$^2$, Xiu-Ding Lin$^1$, Marc J. Feldman$^2$%
, Guang-Can Guo$^1 $}
\address{$^1$Key Laboratory of Quantum Information,\\
University of Science and Technology of China, Chinese Academy of Sciences,\\
Hefei, Anhui 230026, China\\
$^2$Superconducting Electronics Lab, Electrical and Computer Engineering\\
Department,\\
University of Rochester, Rochester, NY 14623, USA }
\maketitle

\begin{abstract}
We present a general framework to study nondistortion quantum interrogation
which preserves the internal state of the quantum object being detected. We
obtain the necessary and sufficient condition for successful performing
nondistortion interrogation for unknown quantum object when the interaction
between the probe system and the detected system takes place only once. When
the probe system and interrogation process have been limited we develop a
mathematical frame to determine whether it is possible to realize NQI
processes only relying on the choice of the original probe state. We also
consider NQI process in iterative cases. A sufficient criterion for NQI is
obtained. \newline
PACS numbers: 03.65.Ta, 42.50.Ct, 03.67.-a
\end{abstract}
\newpage

\section{Introduction}

One of the counterintuitive effects of quantum mechnics is the ``negative
result measurement'', i.e. the nonobservance of a result represents
additional information and hence modifies the wave function. This notion is
first prescribed in 1960 by Renninger\cite{Renninger}, and later Dicke who
analyzed the change of an atomic wave function by the nonscattering of a
photon\cite{Dicke}. In 1993, Elitzur and Vaidman proposed the novel concept
of interaction-free measurements (IFMs), in which the presence of an
absorbing object in a Mach-Zehnder interferometer can be inferred without
apparent interaction with the probe photon\cite{E.V.}. In recent years, EV
IFMs led to numerous investigations and several experiments have been
performed\cite{Kwiat1,Tsegaye,Karlsson,Kwiat2,Rudolph,Geszti,Jang,Mitchison1}%
. In the original EV scheme, the measurement is interaction-free at most
half of the time. However, when combined with quantun zeno effect IFMs can
in principle be done with unity efficiency, in an asympototic sense\cite
{Kwiat1,Kwiat2}.

Aside from their conceptual significance, IFMs are of obvious application
interest too. As emphasized by Vaidman, the paradoxical feature of the EV
IFM is that it obtains information about a region in space without anything
coming in, out, or through this place\cite{Vaidman}. This feature allows
people to monitor radiation sensitive objects without exciting them, such as
the quantum nondemolition measurement of the ground state atom number\cite
{Braginsky}, non-invasive testing of materials\cite{Krenn1,Kent},
localization of atom beams without physical interaction\cite{Krenn},
interaction-free imaging\cite{White} and so on. People also applied the idea
of IFMs in quantum information science, such as quantum cryptography\cite
{Guo} and counterfactual computation\cite{Mitchison2}.

In the field of quantum information science, a physical system may be
regarded as an information carrier if it can be prepared in some
distinguishable quantum states. The advantage of processing information
quantum mechanically is that, contrary to the classical case, information
can be stored in quantum superposition states. However the quantum
superposition is quite fragile. It is interesting to ask whether the similar
treatments on IFMs can be applied in the nondemolition interrogation of the
internal state of quantum object. Suppose there is a black box, in which a
quantum object characterized by its quantum superposition can be persent (or
absent). Our task is to determine if the quantum object in it, without
disturbing the internal state of the object. We may call such an
interrogation a nondistortion quantum interrogation (abbreviated NQI).
Unfortunately, IFMs in general are not internal state preserving
measurements. Since quantum superpositon of the object is subject to
measurement dependent decoherence. Obtaining of ``which way'' information
will cause an unavoidable change of the internal state of the detected
object, even though the measurement is seemingly ``interaction free''.

In this paper we will concentrate on some fundamental limits for the NQI
process, under some definite physical conditions. In Section II, we present
general physical assumptions that we use to formulate the NQI process.
Section III is devoted to single-shot NQI process, in which the interaction
between the detected object and the probe system takes place only once. We
provide the necessary and sufficient condition for successful performing a
single-shot nondistortion interrogation of an unknown quantum state. We also
make a primary attempt to test this criterion when the probe system and the
interaction are restricted. We find that in some simple cases it reduces to
a solvable problem. In Section IV we discuss iterative NQI processes. A
sufficient criterion for Zeno type NQI processes is obtained. We conclude in
Section V by summarizing all the results.

\section{Fundamental Framework of NQI}

``Black box'' problems have attracted people in many contexts. The aim of
the nondistortion quantum interrogation is to find out whether there is a
quantum object in a black box ( a region in space ) without disturbing the
internal state of the object, even if its original quantum state is unknown.

According to Von Neumann's treatments of quantum measurement, any observable
can be coupled to a pointer. The information about the observable is
obtained by detecting the pointer\cite{Preskill}. All the possible
measurement results about this pointer can be described by a set of
orthogonal projectors $\{P_a\}$ in Hilbert space which satisfy 
\begin{equation}
P_a=P_a^{+},P_aP_b=\delta _{ab}P_a,\sum_aP_a=1.  \label{e2.1}
\end{equation}
A Von Neumann measurement will take the probed pure state $\left| \Phi
\right\rangle \left\langle \Phi \right| $ to $\frac{P_a\left| \Phi
\right\rangle \left\langle \Phi \right| P_a}{\left\langle \Phi \right|
P_a\left| \Phi \right\rangle }$ with probability $\Pr (a)=\left\langle \Phi
\right| P_a\left| \Phi \right\rangle $, if the final detected pointer is in
the $P_a$. The original quantum state will be modified by the measurement
process unless the state$\left| \Phi \right\rangle $ is the eigenstate of
the projector $P_a$. Obviously, Von Neumann measurement is not an effiective
way to perform nondistortion interrogation, especially when the quantum
state in the black box is unknown to us.

For the purpose of NQI, we use a probe wave function $\left| \Psi
_{probe}\right\rangle =\alpha \left| \Psi _r\right\rangle +\beta \left| \Psi
_d\right\rangle $. We let part of it $(\left| \Psi _d\right\rangle )$ go
through the black box and interact with the object in the box. After the
interaction $\left| \Psi _r\right\rangle $ and $\left| \Psi _d\right\rangle $
are recombined and a measurement is done on the probe wave function,
possibly after a unitary operation on it (see Fig.1). 
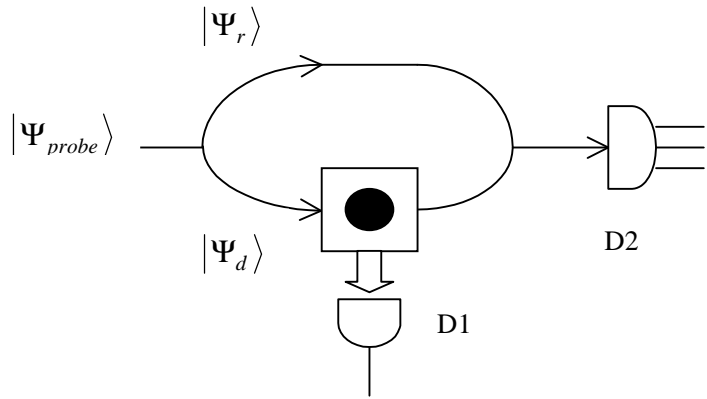
\begin{figure}[h]
\caption{Schematic of the nondistortion interrogation of the quantum object
in the black box. One branch of the probe wave function ($|\Psi _d\rangle $)
goes through the black box and the other branch $|\Psi _r\rangle $ is under
free evolution. The detector $D_1$ registers decay signals from the black
box. If $D_1$ does not fire the two branches are recombined and a Von
Neumann measurement is done on the probe wave function. The final
measurement outcomes will be recorded by $D_2$. }
\label{fig:figure1}
\end{figure}
The above process is a bit similar to the case described by EV scheme.
Indeed, it will be shown that EV IFM is a special case of our ''black box
problem''. While, to perform the NQI in the general case some more exquisite
designs need to be considered. To formulate general NQI processes we make
the following assumptions:

a) $S$, the quantum system under interrogation (the quantum object in the
black box), is a metastable system whose Hilbert space is denoted as $H_S$
with dimension $n$.

b) The Hilbert space of the probe system $D$ , $H_D$ , is composed of two
orthogonal subspaces $H_r$ and $H_d$: $H_D=H_r\bigoplus H_d$. Here, $H_r$
and $H_d$ are the Hilbert spaces of $|\Psi _r\rangle $ and $|\Psi _d\rangle $
respectively and the dimension of the space $H_d$ is $m$. We let $|\Psi
_d\rangle $ go through the black box and interact with S (if the object is
in the box), while $|\Psi _r\rangle $ is under free evolution.

c) The interaction between $D$ and $S$ is governed by a unitary operator. We
assume the time of the interaction is some known $t$. When the interaction
is over, any state driven out of space $H_S$ will quickly decay in an
irreversible way to some stable ground state $|g\rangle $ which is out of
space $H_S$. The decay signal will be registered by some properly arranged
sensitive detectors, if the decay event actually happens.

In assumption a) we note that the physical property of the interrogated
quantum system should be known to us although the original internal state of
this system may be unknown. More exactly, a NQI only refers to an
interrogation for a definite Hilbert space. In assumption b), the
introduction of $H_r$ might appear to be redundant since it does not
interact with $S$. But it is actually vital for the purpose of NQI, because
the interaction between $|\Psi _d\rangle $ and $|\Psi _S\rangle $ changes
the interference between $|\Psi _r\rangle $ and $|\Psi _d\rangle $ which
provides information on what is in the black box. Assumption c) can be named
as ``dissipation postulation'', which makes sure that the state out of the
space $H_S$ can not travel back into the original space $H_S$ after the
interaction between the probe system and the detected system. We assume that
the interrogation process will drive the initial sytem $S$ into an extended
space $H_S\oplus H_S^{\bot }$, while the states in the space $H_S^{\bot }$
will rapidly decay into a stable state $\left| g\right\rangle $ ( which is
out of the space $H_S\oplus H_S^{\bot }$) in an irreversible way. Based on
the above assumptions, the following major steps are followed to find out if
there is an object in the black box ( see Fig.1):

i) Let the probe wave function $|\Psi _d\rangle $ go through the black box
and interact with the quantum object if it is in the box.

ii) The decay signal detectors are used to register any decay event. If a
decay is detected, we conclude that the quantum object was in the black box
but its initial state is destroyed. Otherwise, go to the next step. This is
equivalent to a partial projection measurement on the whole system $\varrho
_{tot}$: 
\begin{equation}
\varrho _{out}=P\rho _{tot}P+P_{\bot }\varrho _{tot}P_{\bot }.  \label{e2.2}
\end{equation}
where the operators $P$ and $P_{\bot }$ refer to unity operators of Hilbert
space $H_S\bigotimes H_D$ and its complementary space $\overline{%
H_S\bigotimes H_D}$ respectively.

iii)Perform a Von Neumann measurement on the probe wave function (in $H_D$).
Possible measurement outcomes are characterized by some projectors in an
orthogonal projector set $O$.

In step iii), we keep in mind that if nothing is in the black box the probe
will be in some definite final state corresponding to the free evolution of
the initial state. We designate the projector to that state as $P_e$. If a
successful interrogation of the object can be done, the probe will end up in
some different state. For the consideration of universality we require that
the probability of successful nondistortion interrogations of the quantum
object be independent on its (unknown) initial state.

We approximately divide the nondistortion interrogation into two types:
single-shot NQI and iterative NQI according to how many times the
interaction happens within an interrogation process. A scheme of single-shot
NQI will follow the above steps i)-iii). In iterative NQI process, step i)
and ii) are repeated. After a certain number of interactions the final
measurement (step iii)) will be executed. Of course, some suitable
operations on the probe wave function are allowed in between the iterations.
Some results about two types of NQI processes are shown in Section III and
IV.

\section{Single-shot NQI process}

Let the initial state of the probe wave function be $\left| \Psi
_{probe}\right\rangle =\alpha \left| \Psi _r\right\rangle +\beta \left| \Psi
_d\right\rangle $, where $\left| \Psi _r\right\rangle \in H_r$ and $\left|
\Psi _d\right\rangle \in H_d$. When the probe passes through the black box
the time evolution of the whole system is as follows: 
\begin{equation}
\left| \Psi _{probe}\right\rangle \left| \Psi _S\right\rangle \rightarrow
\alpha e^{-i/\hbar (H^S+H^D)t}\left| \Psi _r\right\rangle \left| \Psi
_S\right\rangle +\beta e^{-i/\hbar \int_0^t(H^S+H^D+H^I)dt^{\prime }}\left|
\Psi _d\right\rangle \left| \Psi _S\right\rangle   \label{e3.1}
\end{equation}
where $H^S$ and $H^D$ are the free Hamiltonian of the systems $S$ and $D$
respectively, and $H^I$ characterizes the interaction between the two
systems. If there is nothing in the black box the interaction Hamiltonian $%
H^I$ vanishes and the above process reduces to the free evolution of two
separate systems. In contrast, when the black box is occupied with the
quantum object, $\left| \Psi _S\right\rangle $ may be driven into a larger
space $H_S\oplus H_S^{\bot }$ due to the interaction Hamiltonian $H^I$. In
view of the assumption c) a decay could happen with certain probability.
While, if no decay signal is detected, the quantum state of the whole system
is collapsed into the following (a projection on $H_S\otimes H_D$): 
\begin{eqnarray}
&&\alpha e^{-i/\hbar (H^S+H^D)t}\left| \Psi _r\right\rangle \left| \Psi
_S\right\rangle +\beta (I_d\otimes I_S)e^{-i/\hbar
\int_0^t(H^S+H^D+H^I)dt^{\prime }}\left| \Psi _d\right\rangle \left| \Psi
_S\right\rangle   \nonumber \\
&=&\alpha \left| \Psi _r^{\prime }\right\rangle \left| \Psi _S^{\prime
}\right\rangle +\beta (I_d\otimes I_S)e^{-i/\hbar
\int_0^t(H^S+H^D+H^I)dt^{\prime }}e^{i/\hbar (H^S+H^D)t}(I_d\otimes
I_S)\left| \Psi _d^{\prime }\right\rangle \left| \Psi _S^{\prime
}\right\rangle   \label{e3.2}
\end{eqnarray}
where $I_d$ and $I_S$ are the unity operators in the Hilbert spaces $H_d$
and $H_S$, $\left| \Psi _{r(d)}^{\prime }\right\rangle =e^{-i/\hbar
H^Dt}\left| \Psi _{r(d)}\right\rangle $ and $\left| \Psi _S^{\prime
}\right\rangle =e^{-i/\hbar H^St}\left| \Psi _S\right\rangle $. To simplify
the future calculations, the above wave function is unnormalized. In Eq(\ref
{e3.2}), we see that the evolution of the system is fully specified by the
following interrogation operator $D$: 
\begin{equation}
D=(I_d\otimes I_S)e^{-i/\hbar \int_0^t(H^S+H^D+H^I)dt^{\prime }}e^{i/\hbar
(H^S+H^D)t}(I_d\otimes I_S).  \label{e3.3}
\end{equation}

\subsection{necessary and sufficient condition for single-shot NQI}

In the subsection we will review some results in ref\cite{Zhou}.

To obtain the necessary and sufficient condition for single-shot NQI the
following lemma is necessary.

Lemma1: The necessary condition that a single-shot NQI can be done is that
there exist a pair of vectors $|\chi \rangle ,\left| \Psi _d^{\prime
}\right\rangle \in H_d$ which satisfy $\left\langle \chi \right| D\left|
\Psi _d^{\prime }\right\rangle =cI_S$ , where $\left| c\right| \leq 1$.

Proof: If the black box is empty no interaction takes place in it. Thus, the
probe ends up in the state $\alpha |\Psi _r^{\prime }\rangle +\beta |\Psi
_d^{\prime }\rangle $. In term of the promise in above section, this process
will give rise to an exact detecting outcome $P_e=(\alpha |\Psi _r^{\prime
}\rangle +\beta |\Psi _d^{\prime }\rangle )(\alpha ^{*}\langle \Psi
_r^{\prime }|+\beta ^{*}\langle \Psi _d^{\prime }|)\bigotimes I_S$. On the
contrary, the evolution of the probe wave function will be modified when a
quantum object occupies the black box. To make sure that a successful NQI
can be done, there must exist a projector $P_I=|\Psi _I\rangle \langle \Psi
_I|\bigotimes I_S$ ( orthogonal to $P_e$ ) in the set $O$ which corresponds
to the outcome component of NQI with nonzero probability $|\Delta |^2$. This
relation can be embodied in the following equations: 
\begin{equation}
P_IP_e=0  \label{e3.4}
\end{equation}
\begin{equation}
P_I(\alpha |\Psi _r^{\prime }\rangle |\Psi _S^{\prime }\rangle +\beta D|\Psi
_d^{\prime }\rangle |\Psi _S^{\prime }\rangle )=\Delta |\Psi _I\rangle |\Psi
_S^{\prime }\rangle   \label{e3.5}
\end{equation}
where $\left| \Psi _I\right\rangle \in H_D$ is some normalized vector of the
probe. Relying on the Eqs(\ref{e3.4},\ref{e3.5}), we may obtain: 
\begin{equation}
\left\langle \Psi _I\right| D\left| \Psi _d^{\prime }\right\rangle \left|
\Psi _S^{\prime }\right\rangle =(\frac \Delta \beta +\left\langle \Psi
_I|\Psi _d^{\prime }\right\rangle )\left| \Psi _S^{\prime }\right\rangle 
\label{e3.6}
\end{equation}
In terms of the defination of the operator $D$ the vector $D\left| \Psi
_d^{\prime }\right\rangle \left| \Psi _S^{\prime }\right\rangle $ is a
vector in Hilbert space $H_d\otimes H_S$. Thereby, setting a wave vector $%
\left| \chi \right\rangle =I_d\left| \Psi _I\right\rangle $, which is the
projection of $|\Psi _I\rangle $ on $H_d$, we find $\left\langle \Psi
_I\right| D\left| \Psi _d^{\prime }\right\rangle \left| \Psi _S^{\prime
}\right\rangle =\left\langle \chi \right| D\left| \Psi _d^{\prime
}\right\rangle \left| \Psi _S^{\prime }\right\rangle $. Similarly, we also
have the relation $\left\langle \Psi _I|\Psi _d^{\prime }\right\rangle
\left| \Psi _S^{\prime }\right\rangle =\left\langle \chi |\Psi _d^{\prime
}\right\rangle \left| \Psi _S^{\prime }\right\rangle $. Since $\Delta $ and $%
\beta $ are nonzero we deduce that $\left\| \left| \chi \right\rangle
\right\| \neq 0$. Hence, the Eq(\ref{e3.6}) can be rewritten as: 
\begin{equation}
\left\langle \chi \right| D\left| \Psi _d^{\prime }\right\rangle \left| \Psi
_S^{\prime }\right\rangle =(\frac \Delta \beta +\left\langle \chi |\Psi
_d^{\prime }\right\rangle )\left| \Psi _S^{\prime }\right\rangle =c\left|
\Psi _S^{\prime }\right\rangle   \label{e3.7}
\end{equation}
where $c=\frac \Delta \beta +\left\langle \chi |\Psi _d^{\prime
}\right\rangle $. Since $\left| \Psi _S^{\prime }\right\rangle $ is an
arbitrary vector in the space $H_S$ the following must be satisfied: 
\begin{equation}
\left\langle \chi \right| D\left| \Psi _d^{\prime }\right\rangle =cI_S.
\label{e3.8}
\end{equation}
We thus complete our proof.

Once the necessary condition (\ref{e3.8}) is satisfied the interrogation
operator $D$ will modify the wavefunction component $\left| \Psi _d^{\prime
}\right\rangle \left| \Psi _S^{\prime }\right\rangle $ in the following way: 
\begin{equation}
D\left| \Psi _d^{\prime }\right\rangle \left| \Psi _S^{\prime }\right\rangle
=c\left| \chi \right\rangle \left| \Psi _S^{\prime }\right\rangle
+\sum_{j=1}^{m-1}\left| \widetilde{\chi _j}\right\rangle \left| \widetilde{%
m_{S(j)}}\right\rangle .  \label{e3.91}
\end{equation}
Here, $\left| \widetilde{m_{S(j)}}\right\rangle =\left\langle \widetilde{%
\chi _j}\right| D\left| \Psi _d^{\prime }\right\rangle \left| \Psi
_S^{\prime }\right\rangle $ and the vectors $\{\left| \chi \right\rangle
,\left| \widetilde{\chi _j}\right\rangle ;j=1,...,m-1\}$ form the orthogonal
complete bases in Hilbert space $H_d$.

As we know, the decomposition of a pure quantum state is roughly dependent
on the structure of its Hilbert space. However, it should be noted that the
representation of Eq(\ref{e3.91}) is not enough ``compact''. It seems too
extravagant for decomposing the state $D\left| \Psi _d^{\prime
}\right\rangle \left| \Psi _S^{\prime }\right\rangle $ in the Hilbert space $%
H_d\otimes H_S$. Whether we can choose a more compact subspace in the space $%
H_d\otimes H_S$ to achieve a full decomposition for it? This point is quite
vital for obtaining the necessary and sufficient condition for the
single-shot NQI. To express the Eq(\ref{e3.91}) in a more compact way the
following steps should be done. We may define a set of operators $Q^{\left(
i\right) }=Tr_S\left[ D\left| \Psi _d^{\prime }\right\rangle \left|
i\right\rangle \left\langle i\right| \left\langle \Psi _d^{\prime }\right|
D^{+}\right] $ if the Eq(\ref{e3.8}) holds. Here, $\{\left| i\right\rangle
,i=1,...,n\}$ is a set of orthogonal bases in the Hilbert space $H_S$. In
the Hilbert space $H_d$, the kernel operator of the operator $Q^{\left(
i\right) }$ is denoted as $K_i$. Thus, the operator $K_i\otimes I_S$ can be
seen as the annihilation operator for the component $D\left| \Psi _d^{\prime
}\right\rangle \left| i\right\rangle $: 
\begin{equation}
K_i\otimes I_S(D\left| \Psi _d^{\prime }\right\rangle \left| i\right\rangle
)=0.  \label{e3.92}
\end{equation}
If we set the intersection of all the $n$ kernel operators as $%
K=\bigcap_{i=1}^nK_i$ the operator $K\otimes I_S$ will annihilate the
quantum state $D\left| \Psi _d^{\prime }\right\rangle \left| \Psi _S^{\prime
}\right\rangle $. Because 
\begin{equation}
K\otimes I_S(D\left| \Psi _d^{\prime }\right\rangle \left| \Psi _S^{\prime
}\right\rangle )=\sum_{i=1}^nc_i^SK\otimes I_S(D\left| \Psi _d^{\prime
}\right\rangle \left| i\right\rangle )=0  \label{e3.93}
\end{equation}
where $\left| \Psi _S^{\prime }\right\rangle =\sum_{i=1}^nc_i^S\left|
i\right\rangle $. Therefore, a more compact space for decomposing the
quantum state $D\left| \Psi _d^{\prime }\right\rangle \left| \Psi _S^{\prime
}\right\rangle $ is $H_{\overline{K}}\otimes H_S$. Here, $H_{\overline{K}}$
is the complementary space of kernel space $H_K$ in $H_d$. We denote the
dimension of the space $H_{\overline{K}}$ by $l(l\leq m)$. When we pick up
some set of orthonormal states $\{\left| \chi \right\rangle ,\left| \chi
_1\right\rangle ,...,\left| \chi _{l-1}\right\rangle \}$ spanning the space $%
\overline{K}$. Then, Eq(\ref{e3.91}) can be rewritten as follows: 
\begin{eqnarray}
D\left| \Psi _d^{\prime }\right\rangle \left| \Psi _S^{\prime }\right\rangle
&=&c\left| \chi \right\rangle \left| \Psi _S^{\prime }\right\rangle
+\sum_{j=1}^{l-1}\left| \chi _j\right\rangle \left| m_{S(j)}\right\rangle 
\nonumber \\
\left| m_{S(j)}\right\rangle &=&\left\langle \chi _j\right| D\left| \Psi
_d^{\prime }\right\rangle \left| \Psi _S^{\prime }\right\rangle .
\label{e3.9}
\end{eqnarray}
We may outline the above treatments into the following lemma.

Lemma2: Eq(\ref{e3.9}) is the equivalent representation of the Eq(\ref{e3.8}%
).

In light of the above preparations we may provide our main theroem.

Theorem 1: The necessary and sufficient condition for the single-shot NQI is
that Eq(\ref{e3.9}) holds and $\left| \Psi _d^{\prime }\right\rangle
-c\left| \chi \right\rangle $ is linearly independent of the state set $%
\{\left| \chi _j\right\rangle ;j=1,...,l-1\}$.

Proof: Lemma 1 and 2 show that Eq(\ref{e3.9}) is the necessary condition for
single-shot NQI. If Eq(\ref{e3.9}) holds, in Hilbert space $H_S\bigotimes
H_D $ the final state of the whole system will be: 
\begin{equation}
\left| \Psi _{probe}\right\rangle \left| \Psi _S\right\rangle \rightarrow
\alpha \left| \Psi _r^{\prime }\right\rangle \left| \Psi _S^{\prime
}\right\rangle +\beta c\left| \chi \right\rangle \left| \Psi _S^{\prime
}\right\rangle +\beta \sum_{j=1}^{l-1}\left| \chi _j\right\rangle \left|
m_{S(j)}\right\rangle .  \label{e3.10}
\end{equation}
Reconsidering Eqs. (\ref{e3.4}) and (\ref{e3.5}), we may attain the
following relations: 
\begin{equation}
\left\langle \Psi _I\right| \left( \alpha \left| \Psi _r^{\prime
}\right\rangle +\beta \left| \Psi _d^{\prime }\right\rangle \right) =0
\label{e3.11}
\end{equation}
\begin{equation}
\left\langle \Psi _I|\chi _j\right\rangle =0  \label{e3.12}
\end{equation}
\begin{equation}
\left\langle \Psi _I\right| \left( \alpha \left| \Psi _r^{\prime
}\right\rangle +c\beta \left| \chi \right\rangle \right) =\Delta .
\label{e3.13}
\end{equation}
Subtracting Eq(\ref{e3.13}) from Eq(\ref{e3.11}) we obtain 
\begin{equation}
\left\langle \Psi _I\right| \beta \left( \left| \Psi _d^{\prime
}\right\rangle -c\left| \chi \right\rangle \right) =-\Delta .  \label{e3.14}
\end{equation}
$\Delta \neq 0$ requires that $\left| \Psi _d^{\prime }\right\rangle
-c\left| \chi \right\rangle $ be linearly independent of the set of vectors $%
\{\left| \chi _j\right\rangle ;j=1,...,l-1\}$. We thus confirm the necessity
of the criterion.

We may further prove the converse by constructing a projector $P_I$
satisfying Eq(\ref{e3.4}) and Eq(\ref{e3.5}). This procedure can appeal to
the Schmidt orthogonalization method. First we define the state set $N$
consisting of $l+1$ normalized vectors $\{\alpha \left| \Psi _r^{\prime
}\right\rangle +\beta \left| \Psi _d^{\prime }\right\rangle ,\gamma \left(
\alpha \left| \Psi _r^{\prime }\right\rangle +c\beta \left| \chi
\right\rangle \right) ,\left| \chi _j\right\rangle ;j=1,...,l-1\}$, where $%
\gamma =\frac 1{\left\| \alpha \left| \Psi _r^{\prime }\right\rangle +c\beta
\left| \chi \right\rangle \right\| }$ is the normalization coefficient for $%
\alpha \left| \Psi _r^{\prime }\right\rangle +c\beta \left| \chi
\right\rangle $. We assume that $\alpha \neq 0$. Since $\left| \chi
\right\rangle ,\left| \chi _1\right\rangle ,...\left| \chi
_{l-1}\right\rangle $ and $\left| \Psi _r^{\prime }\right\rangle $ are
orthogonal to each other and $\left| \Psi _d^{\prime }\right\rangle -c\left|
\chi \right\rangle $ is linearly independent of $\{\left| \chi
_j\right\rangle ;j=1,...,l-1\}$, we may deduce that all vectors in the state
set $N$ are linearly independent. To construct an orthonormal set out of $N$%
, we let the first $l-1$ vectors be $\{\left| \chi _j\right\rangle
;j=1,...,l-1\}$. We then calculate the $l$th vector using $\alpha \left|
\Psi _r^{\prime }\right\rangle +\beta \left| \Psi _d^{\prime }\right\rangle $%
: $\left| \widetilde{\Psi }\right\rangle =\gamma ^{\prime }(\alpha |\Psi
_r^{\prime }\rangle +\beta \Psi _d^{\prime }\rangle -\Sigma
_{i=1}^{l-1}\left\langle \chi _i\right| (\alpha |\Psi _r^{\prime }\rangle
+\beta \Psi _d^{\prime }\rangle )\left| \chi _i\right\rangle )$, where $%
\gamma ^{\prime }$ is the normalization coefficient. Similarly, the last
vector is $\left| \Psi _I\right\rangle =\gamma ^{\prime \prime }\left(
\gamma \left( \alpha \left| \Psi _r^{\prime }\right\rangle +c\beta \left|
\chi \right\rangle \right) -\left\langle \widetilde{\Psi }\right| \gamma
\left( \alpha \left| \Psi _r^{\prime }\right\rangle +c\beta \left| \chi
\right\rangle \right) \left| \widetilde{\Psi }\right\rangle \right) $with
the normalization coefficient $\gamma ^{\prime \prime }$. It is then obvious
that the projector $P_I=|\Psi _I\rangle \langle \Psi _I|\bigotimes I_S$
satisfies Eqs(\ref{e3.4}) and (\ref{e3.5}). In addition, once the
coefficient $\alpha $ is fixed the projector $P_I$ obtained by using Schmidt
orthogonalization will maximize the success probobility $\left| \Delta
\right| ^2$ \cite{Zhou}. We thus complete our proof.

From our proof it is clear that $|\Psi _r^{\prime }\rangle $ is not
redundant, even though it does not interact with the object. If $\alpha =0$
the capability of performing an NQI will be severely limited. For instance,
if $c$ is zero in Eq(\ref{e3.9}) (This corresponds to an IFM of an opaque
object in which the probe wave function is blocked by the absorbing object),
no nondistortion interrogation of the object can be done without the
introduction of $|\Psi _r^{\prime }\rangle $. We may elucidate the novel NQI
phenomena in the sense of quantum interference. The quantum object in the
black box can be seen as a scattering object corresponding to the probe
wave. The scattering process between the object and the probe wave makes
each scattering wave component entangle different evolution of the object.
If all the information on the evolution of the whole composite system is
known it possibly allows us to choose a proper probe wave such that a
successful scattering wave component is produced. For the purpose of NQI\
this component gets entangled with the free evolution of the object. Now, it
seems clearly for the meaning of the above theorem. Once the necessary and
sufficient condition for single-shot NQI holds the form of this successful
probe wave component $|\Psi _I\rangle $ can be obtained by using Schmidt
orthogonalization steps outlined in the proof of theorem 1.

We note that the vectors $\left| \Psi _d^{\prime }\right\rangle $ and $%
\left| \chi \right\rangle $ satisfying Eq(\ref{e3.8}) may not be unique.
However, once $\left| \chi \right\rangle $ and $\left| \Psi _d^{\prime
}\right\rangle $ are chosen we may obtain the optimal success probability
for these two vectors, by following the steps outlined in the proof of
theorem 1. If the initial state of the probe is $\alpha \left| \Psi
_r\right\rangle +\beta \left| \Psi _d\right\rangle $, the success
probability is 
\begin{equation}
\Pr ob(\alpha )=\left| \left\langle \Psi _I\right| (\alpha \left| \Psi
_r^{\prime }\right\rangle +c\beta \left| \chi \right\rangle )\right| ^2.
\label{e3.15}
\end{equation}
Therefore we have the following corollary.

Corollary 1: Under the condition that the wave function $\left| \Psi
_d^{\prime }\right\rangle $ and $\left| \chi \right\rangle $ are given the
optimal success probability of the NQI is as follows: 
\begin{equation}
P_{opt}=\max_{\left| \alpha \right| \in \left[ 0,1\right) }\Pr ob(\alpha ).
\label{e3.16}
\end{equation}

\subsection{Verifying the criterion of NQI}

In the former context we have pointed out the operator $D$ fully
characterizes the interaction between two systems and the dissipation
process. In this subsection, starting from the operator $D$ we will devote
ourself to explore the existance of the pair of vectors $\left| \Psi
_d^{\prime }\right\rangle $ and $\left| \chi \right\rangle $ satisfying the
criterion of NQI.

The operator $D$ can be written as: 
\begin{equation}
D=\sum_{i,j;k,l}d_{i,j;k,l}\left| i_d\right\rangle \left| k_S\right\rangle
\left\langle l_S\right| \left\langle j_d\right| .  \label{e3.17}
\end{equation}
For the simplicity of discussion we will omit the subscripts of the vectors.
Unless pointed out otherwise $\left| i\right\rangle \left( \left|
j\right\rangle \right) $ will indicate the vectors in the Hilbert Space $H_d$
and $\left| k\right\rangle \left( \left| l\right\rangle \right) $ will
indicate the vectors in the Hilbert Space $H_S$.

Once a successful NQI\ process can be put in practice the operator $D$, as a
subblock in a unitary matrix( see Eq(\ref{e3.3})), satisfies the following
relation: 
\begin{equation}
\left\langle \chi \right| D\left| \Psi _d^{\prime }\right\rangle =cI_S.
\label{e3.18}
\end{equation}
We set $\left| \chi \right\rangle =\sum_{i=1}^ma_i^{*}\left| i\right\rangle $
and $\left| \Psi _d^{\prime }\right\rangle =\sum_{i=1}^mb_i\left|
i\right\rangle $ the relation(\ref{e3.18}) can be parameterized as: 
\begin{equation}
\sum_{i,j}a_id_{i,j;k,l}b_j=c\delta _{kl},(k,l=1,2,...,n)  \label{e3.19}
\end{equation}
Furthermore, we may define $\sum_jd_{i,j;k,l}b_j$ as a matrix $\Re _{i;k,l}$
with a set of parameters $\left\{ b_j\right\} $. Thus, whether the equation (%
\ref{e3.18}) holds is equivalent to whether there are a set of parameters $%
\left\{ a_i\right\} $satisfying the following equation: 
\begin{equation}
\sum_ia_i\Re _{i;k,l}=c\delta _{kl},(k,l=1,2,...,n)  \label{e3.20}
\end{equation}
Since the above equations are linear, in principle, it is possible to
identify the existence of the unknown parameters $\left\{ a_i\right\} $
satisfying the equation set(\ref{e3.20}).

Theorem 2: the necessary and sufficient condition for the existence of the
solution is that the matrix $(\Re )_{m\times n^2}$ and its augmented matrix $%
(\widetilde{\Re })_{(m+1)\times n^2}$ has the same rank(see ref\cite{chen}): 
\begin{equation}
rank(\Re )=rank(\widetilde{\Re })  \label{e3.20a}
\end{equation}
where the augmented matrix $(\widetilde{\Re })$ is 
\begin{equation}
\widetilde{\Re }_{i;k,l}=\Re _{i;k,l}\left( i\leq m\right) ,\widetilde{\Re }%
_{m+1;k,l}=c\delta _{kl}.  \label{e3.20b}
\end{equation}

Unfortunately, since the matrix $\Re _{i;k,l}$ includes the uncertain
parameters $\left\{ b_j\right\} _{j=1}^m$, how to develop a general method
to determine the existence of the solution of the equation set (\ref{e3.20})
remains unknown. Numerically, we may simply try all possible $b_j$'s $\left(
\sum_j\left| b_j\right| ^2=1\right) $ and $c$ to see if Eq(\ref{e3.20a}) is
satisfied. When the total number of the equations in (\ref{e3.20}) is not
larger than $m$ it is a readily solvable problem. In principle, if the
solutions exist, it can be formally described as: 
\begin{equation}
a_i=a_i(b_1,b_2,...,b_m,c);i=1,...,m.  \label{e3.20c}
\end{equation}
Furthermore, according to the steps shown in the above context we may test
the criterion of the NQI. The solutions of the equation set (\ref{e3.20})
may not be unique, that is, for different $b_i$'s and $c$ we may get
different $a_i$'s. In practice, we may choose the set of solutions which
maximize the success probability. Here, Let us consider a concrete example.

As in Fig. 2, the model we consider is a multi-level atom. The atom is
prepared in an arbitrary superposition of the two degenerate metastable
states $\left| m+\right\rangle $ and $\left| m-\right\rangle $. By obsorbing
a $+\left( -\right) $ circularly polarized photon with the frequency $\omega 
$ the atom can make a resonant transition from $\left| m+\right\rangle
\left( \left| m-\right\rangle \right) $ to the corresponding exicted state $%
\left| e+\right\rangle $ $(\left| e-\right\rangle )$. Here, we may view the
transition between $\left| m+\right\rangle \left( \left| m-\right\rangle
\right) $ and $\left| e+\right\rangle $ $(\left| e-\right\rangle )$ caused
by the $+\left( -\right) $ circularly polarized photon as a resonant
Jaynes-Cummings model. Clearly, there are no correlations between the two
processes. We assume that the atom will decay rapidly from the excited state 
$\left| e+\right\rangle \left( \left| e-\right\rangle \right) $ to the
ground state $\left| g\right\rangle $ in an irreversible way when the
electromagnetic field with the frequency $\omega $ fades away. At last, the
decay signals can be recorded by some sensitive detectors. To study the
criterion of NQI process we start with the Hamiltonian of the whole system: 
\begin{eqnarray}
H_{tot} &=&H_{atom}+H_{photon}+H_{interaction}  \nonumber \\
\ &=&\sum_{k=+,-}\left[ \hbar \omega a_k^{+}a_k+\hbar \omega \left|
ek\right\rangle \left\langle ek\right| +\hbar g_k\left( \sigma
_{+}^ka_k+a_k^{+}\sigma _{-}^k\right) \right]  \label{e3.21}
\end{eqnarray}
\begin{figure}[h]
\caption{Level structure of the atom. The atom can make a transition to the
excited state $|e_{+}\rangle $ $(|e_{-}\rangle )$ from $|m_{+}\rangle $ $%
(|m_{-}\rangle )$ by absorbing a $+(-)$ circular polarized photon. It then
decays rapidly to the stable ground state $|g\rangle $.}
\label{fig:figure2}
\end{figure}
where $a_k^{+}$ and $a_k$ are the creation and annihilation operators of the 
$k$ polarized photon. the operators $\sigma _{+}^k$ and $\sigma _{-}^k$
separately refer to $\left| ek\right\rangle \left\langle mk\right| $ and$%
\left| mk\right\rangle \left\langle ek\right| $. Here, we define $\left|
mk\right\rangle $ as the zero point of the energy. To formulate the operator 
$D$ our main task is to obtain the representation of the unitary operator $%
U(t)=e^{-i/\hbar H_{tot}t}$. For that purpose, we may bring in the total
particle number operator $N=\sum_{k=+,-}(\hbar \omega a_k^{+}a_k+\hbar
\omega \left| ek\right\rangle \left\langle ek\right| )$. Since the operator $%
N$ commutes with the Hamiltonian $H_{tot}$, i.e. $\left[ N,H_{tot}\right] =0$%
, the unitary operator $U(t)$ can be decomposed into the direct summands of
a series of subspaces labelled by the eigenvalues of the operator $N$. As a
prior assumption of this NQI process, we assume that only one photon comes
into the system and interacts with the atom prepared in the superposition of
the metastable states. Due to the fact that $(N-\hbar \omega )\left|
one\_photon\right\rangle \left| metastable\_state\right\rangle =0$, during
the whole interaction process, the quantum state of the whole system can not
escape from the eigenspace of the operator $N$ with the eigenvalue $\hbar
\omega $. This subspace is spanned by six basis vectors: $\left|
ek\right\rangle \left| 0\right\rangle $, $\left| mk\right\rangle \left|
k^{\prime }\right\rangle $ $\left( k,k^{\prime }=+,-\right) $. Here, $\left|
k\right\rangle $ and $\left| 0\right\rangle $ refer to the state of single $%
k $ (circularly) polarized photon and the vacuum state respectively. Thus,
in this problem, we need only give the representation of the unitary
operator $U(t)$ in the eigenspace of the operator $N$ labelled by eigenvalue 
$\hbar \omega $. 
\begin{eqnarray}
U^{(\hbar \omega )}(t) &=&e^{-i\omega t}\{\sum_{k=+,-}\left( \cos g_kt\left|
mk\right\rangle \left| k\right\rangle \left\langle k\right| \left\langle
mk\right| -i\sin g_kt\left| ek\right\rangle \left| 0\right\rangle
\left\langle k\right| \left\langle mk\right| \right)  \nonumber \\
&&\ +\sum_{k=+,-}\left( \cos g_kt\left| ek\right\rangle \left|
0\right\rangle \left\langle 0\right| \left\langle ek\right| -i\sin
g_kt\left| mk\right\rangle \left| k\right\rangle \left\langle 0\right|
\left\langle ek\right| \right)  \nonumber \\
&&\ +\left| m-\right\rangle \left| +\right\rangle \left\langle +\right|
\left\langle m-\right| +\left| m+\right\rangle \left| -\right\rangle
\left\langle -\right| \left\langle m+\right| \}.  \label{e3.22}
\end{eqnarray}
Based on the equation(\ref{e3.3}) the operator $D$ can be obtained: 
\begin{equation}
D=\sum_{k=+,-}\cos g_kt\left| mk\right\rangle \left| k\right\rangle
\left\langle k\right| \left\langle mk\right| +\left| m-\right\rangle \left|
+\right\rangle \left\langle +\right| \left\langle m-\right| +\left|
m+\right\rangle \left| -\right\rangle \left\langle -\right| \left\langle
m+\right| .  \label{e3.23}
\end{equation}
We may further deduce that the space $\overline{K}$ is just $H_d$ spanned by
two vectors $\left| +\right\rangle $ and $\left| -\right\rangle $. By
setting $\left| \chi \right\rangle =\sum_{i=+,-}a_i^{*}\left| i\right\rangle 
$ and $\left| \Psi _d^{\prime }\right\rangle =\sum_{i=+,-}b_i\left|
i\right\rangle $ we can rephrase the equation(\ref{e3.20}) as follows: 
\begin{equation}
\left[ 
\begin{array}{cc}
p_{+}b_{+} & b_{-} \\ 
b_{+} & p_{-}b_{-}
\end{array}
\right] \left[ 
\begin{array}{c}
a_{+} \\ 
a_{-}
\end{array}
\right] =\left[ 
\begin{array}{c}
c \\ 
c
\end{array}
\right]  \label{e3.24}
\end{equation}
where $p_{\pm }=\cos g_{\pm }t$. To make sure that there are rational
solutions of $a_{+}$ and $a_{-}$ it is essential that the determinant of the
matrix on the left hand of the above equation is non-zero: $%
(p_{+}p_{-}-1)b_{+}b_{-}\neq 0$ i. e. $p_{+}p_{-}\neq 1$ and $b_{+}b_{-}\neq
0$. The solutions of the equation set(\ref{e3.24}) are that: $a_{+}=\frac{%
\left( p_{-}-1\right) c}{\left( p_{+}p_{-}-1\right) b_{+}}$, $a_{-}=\frac{%
\left( p_{+}-1\right) c}{\left( p_{+}p_{-}-1\right) b_{-}}$. the
normolizaton of the vector $\left| \chi \right\rangle $ provides the limit
of the norm of the constant $c$: $\left| c\right| =\frac{\left|
b_{+}b_{-}\right| \left( 1-p_{+}p_{-}\right) }{\sqrt{\left( p_{-}-1\right)
^2\left| b_{-}\right| ^2+\left( p_{+}-1\right) ^2\left| b_{+}\right| ^2}}$.
For the purpose of simplicity we set $p_{+}=p_{-}=p$. Thus, the above
parameters should be separately reduced to: $a_{+}=\frac c{\left( p+1\right)
b_{+}}$, $a_{-}=\frac c{\left( p+1\right) b_{-}}$, and $\left| c\right|
=\left| b_{+}b_{-}\right| \left( 1+p\right) $. Starting from the state
vector $\left| \chi \right\rangle =\frac{c^{*}}{\left( p+1\right) b_{+}^{*}}%
\left| +\right\rangle +\frac{c^{*}}{\left( p+1\right) b_{-}^{*}}\left|
-\right\rangle $ we can deduce $\left| \chi _1\right\rangle =\frac c{\left(
p+1\right) b_{-}}\left| +\right\rangle -\frac c{\left( p+1\right) b_{+}}%
\left| -\right\rangle $. By comparing $\left| \Psi _d^{\prime }\right\rangle
-c\left| \chi \right\rangle $ with $\left| \chi _1\right\rangle $ we may
find that the two state vectors are linearly independent unless $p=1$. In
other words, for the atomic system as depicted in Fig. 2, the criterion of
NQI\ process can be satisfied in most cases. Therefore, it allows us to
obtain the projector $P_I$ by taking advantage of the steps outlined in
section III A and to devise feasible protocols of nondistortion
interrogation. One such example is given in \cite{Potting}, where $\left|
\chi \right\rangle =\left| \psi _d^{\prime }\right\rangle =\frac 1{\sqrt{2}}%
\left( a_{l,-}^{+}-a_{l,+}^{+}\right) \left| 0\right\rangle $, and the
maximum success probability is $\frac 1{16}$.

\section{Iterative NQI\ process}

People have recently presented several protocols of NQI\cite
{Potting,Xiang1,Xiang2}. However, it is quite difficult to increase the
interrogation efficiency in single-shot NQI\ processes. We proved that the
optimal success probability in the protocol investigated by P\"otting et. al.%
\cite{Potting} can only reach $\frac 1{16}$\cite{Zhou}. By expanding the
Hilbert space of the probe system, the success probability for NQI can be
raised to $\frac 18$ in an advanced scheme\cite{Xiang1}. The design for the
previous two schemes is just in terms of the fundamental steps i)-iii)
described in section II. As we know, in most protocols of the
interaction-free measurements, it is possible to make the probability of
IFMs arbitrarily close to one by taking advantage of the quantum Zeno effect%
\cite{Kwiat1,Tsegaye,Kwiat2,Mitchison1}. Similar treatments can also be
considered in NQI. A primary attempt was presented in ref\cite{Xiang2}.

For the iterative NQI\ processes, a basic consideration is that when the
probe wavefunction comes out of the detected system one makes it come in the
detected system again after an appropriate manipulation on it, instead of
performing the Von Neumann measurement on it. A final measurement can be
done on the probe after the certain number of iterations. When we denote the
quantum manipulation between the loops by the operator $L^{\left( i\right) }$
the final state in $H_D\bigotimes H_S$ is: 
\begin{equation}
\left| \Psi \right\rangle _{non-decay}=\prod_{i=0}^{N-1}\left( L^{\left(
N-i\right) }\left( \left( I_r\otimes I_S\right) \oplus D\right)
U_{free}\right) \left| \Psi _D\right\rangle \left| \Psi _S\right\rangle
\label{e4.1}
\end{equation}
where $U_{free}=e^{-i/\hbar (H^D+H^S)t}$ and N refers to the number of the
iterations. Because we allow to insert the proper operation $L^{\left(
i\right) }$ in between the iterations it becomes hard to obtain the
necessary and sufficient for the iterative NQI processes. Here, we only
consider the sufficient condition for the iterative cases.

Theorem 3: The sufficient condition that an iterative NQI can be done with
certainty is that there exists a vector $|\chi \rangle \in H_d$ which
satisfies $\left\langle \chi \right| D\left| \chi \right\rangle =cI_S$ ,
where $c$ is a known constant and is not equal to one.

Proof: Suppose that the original probe wavefunction is: $\left| \Psi
_D\right\rangle =e^{i/\hbar H^Dt}(\cos \theta \left| \psi _r\right\rangle
+\sin \theta \left| \chi \right\rangle )$. If the condition $\left\langle
\chi \right| D\left| \chi \right\rangle =cI_S\left( c\neq 1\right) $ holds,
for any state vector $\left| \psi _S^{\prime }\right\rangle \in S\left(
H_S\right) $, we have the following relation: 
\begin{equation}
D\left| \chi \right\rangle \left| \psi _S^{\prime }\right\rangle =c\left|
\chi \right\rangle \left| \psi _S^{\prime }\right\rangle
+\sum_{j=1}^{m-1}\left| \chi _j\right\rangle \left| m_{S(j)}\right\rangle .
\label{e4.2}
\end{equation}
Here, $\left\{ \left| \chi _j\right\rangle _{j=1}^{m-1},\left| \chi
\right\rangle \right\} $ form the bases of the Hilbert space $H_d$ and $%
\left| m_{S(j)}\right\rangle =\left\langle \chi _i\right| D\left| \chi
\right\rangle \left| \psi _S^{\prime }\right\rangle $. For the iterative
NQI, the main task is to elaborately devise the operations between the
adjacent loops. Considering the condition(\ref{e4.2}) we may set the
operator $L^{\left( i\right) }$ as follows: 
\begin{equation}
L^{\left( i\right) }=e^{i/\hbar H^Dt}UM  \label{e4.3}
\end{equation}
where $M$ is a projector which projects the probe wavefunction into the
subspace $\left( \left| \psi _r\right\rangle \left\langle \psi _r\right|
+\left| \chi \right\rangle \left\langle \chi \right| \right) \otimes I_S$
with some probability. The unitary operator $U$ has the following form: 
\begin{equation}
U=\left( 
\begin{array}{cc}
\cos \delta & \sin \delta \\ 
-\sin \delta & \cos \delta
\end{array}
\right)  \label{e4.4}
\end{equation}
where the matrix is in the bases$\left| \psi _r\right\rangle $ and $\left|
\chi \right\rangle $. We choose $\delta =\theta -\theta ^{\prime }$, where $%
\cos \theta ^{\prime }=\cos \theta /\sqrt{\cos ^2\theta +c^2\sin ^2\theta }$%
. Thus, if there is a quantum object characterized in the state $\left| \psi
_S\right\rangle $ in the black box, after the probe wavefunction initially
prepared in the state $\left| \Psi _D\right\rangle $ undergoes N iterations,
if we choose N such that $N\delta =\pi /2$, the non-decay quantum fraction
of the whole system will be transformed into: 
\begin{equation}
\left| \Psi \right\rangle _{non-decay}=e^{\frac{iH^Dt}\hbar }\gamma ^N\left(
\cos \theta \left| \psi _r\right\rangle +\sin \theta \left| \chi
\right\rangle \right) e^{-\frac{iH^SNt}\hbar }\left| \psi _S\right\rangle
\label{e4.5}
\end{equation}
where $\gamma =1/\sqrt{\cos ^2\theta +c^2\sin ^2\theta }$. As the final
result, the probe will be end up in the state $e^{\frac{iH^Dt}\hbar }\left(
\cos \theta \left| \psi _r\right\rangle +\sin \theta \left| \chi
\right\rangle \right) $ with the probability $\gamma ^{2N}$, which goes to
unity in the limit of large N and small $\theta $ (see the ref\cite
{Mitchison1}). On the contrary, if the black box is empty the quantum state
of the probe will evolve into: 
\begin{equation}
\left| \Psi \right\rangle =e^{\frac{iH^Dt}\hbar }\left[ \cos (\theta +\pi
/2)\left| \psi _r\right\rangle +\sin (\theta +\pi /2)\left| \chi
\right\rangle \right]  \label{e4.6}
\end{equation}
with certainty. Due to the orthogonality and (asymptotic) certainty of the
two outcomes we may perfectly perform the NQI in the iterative way. This
completes the proof of theroem 3.

\section{Discussion and Conclusion}

In the quantum interrogation, the interaction between the probe wave and the
quantum object modifies the interference among the original probe wave
components. Meanwhile, each of the probe wave components entangles the
corresponding evolution of the scattered object. For the NQI processes,
there must exist a successful probe wave component, which should entangle
with the free evolution of the object, at the same time, be orthogonal to
any other scattering wave components. Quantum measurement theory ensures
that once this component is registered the corresponding quantum branch will
come into reality. In this case, we may not only obtain information on the
location of a quantum object, but also make the evolution of the internal
state of the detected object free from the disturbance from the
interrogation process. Here, we should emphasize that before an
experimentist provides a practical NQI\ scheme all the physical conditions
about the interaction in the black box should be considered carefully. In
addition, As far as some complicated interactions are concerned, it is quite
difficult to test the criterion of NQI. Especially, how to obtain the
sufficient and necessary criterion for the iterative NQI processes is still
an open question.

Nevertheless, NQI may provide some miraculous applications. It should be
noted that the interrogated target is not a quantum state, but a space of
quantum states. In other words, quantum information in Hilbert space $H_S$
will not be contaminated by the probing process. This way to manipulate
quantum objects is of potential application in the recently developed
quantum information science. Since the detected system need not be
restricted to pure states we may also cast our interests on mixed states. As
pointed out by P\"otting et. al \cite{Potting} the nondistortion
interrogation provides a tool to monitor a subsystem in a many-particle
system without destroying the entanglement between the particles.

In conclusion, we have studied the process of NQI under some physical
assumptions. We proved the necessary and sufficient condition for
single-shot NQI process in our formulation. Furthermore, we consider the NQI
processes in iterative cases and obtain a sufficient condition for the
iterative nondistortion quantum interrogation. As a novel method to
manipulate quantum systems NQI may be applied in future quantum information
processing.

Z. W. Zhou and G. C. Guo are funded by National Fundamental Research
Program( 2001CB309300), National Natural Science Foundation of China ( Grant
No. 10204020), the Innovation funds from Chinese Academy of Sciences. Work
of X. Zhou and M. J. Feldman was supported in part by ARO grant
DAAG55-98-1-0367.

\end{document}